\documentstyle[11pt,newpasp,twoside,epsf]{article} 
\def\edcomment#1{\iffalse\marginpar{\raggedright\sl#1\/}\else\relax\fi} 
\reversemarginpar 

\begin{document} 
\title{Stellar Abundances in 
Giant Stars in the Metal-
Rich Globular Cluster
 NGC6528}

\author{Sofia Feltzing} 
\affil{Lund Observatory, Box 43, SE-221 00 Lund} 
\author{Francesca Primas \& Rachel Johnson}
\affil{ESO, Garching \& Santiago} 

\begin{abstract} 
We present the first results of a detailed abundance analysis,
based on VLT observations, 
of giant stars in the very metal-rich globular cluster NGC 6528.
 We will, we hope, be able to 
tie the horizontal branch abundances (see e.g. Carretta et al. 2001) to
those of the more luminous giants (see CMDs). For the very similar 
cluster NGC 6553 studies of different types of stars have yielded
very disparate results. Our first analysis of three of our stars
seem to indicate that indeed the different sorts of stars do show similar
abundances if one homogeneous set of models and parameters are being used.
\end{abstract}

\smallskip
\noindent
{\bf Introduction}
NGC 6528 is perhaps {\it the} most metal-rich globular
cluster known and as such serves as an invaluable template  for
calibrating and understanding  metal-rich stellar populations in
extra-galactic bulges and disks, as well as serving as a template  in
studies of our own Galactic Bulge. So far, observations of this cluster 
have been, however, severely hampered by it being superimposed on and 
physically close to the Galactic Bulge itself. But now, using our second 
epoch HST/WFPC2 observations, we were able to provide a disentangling of 
Bulge and cluster and could therefore propose a detailed 
abundance analysis of {\sl bona fide} cluster members. These would determine 
the metallicity of the cluster (very much debated, values 
between -0.6 to +0.2 dex 
are found in the literature), as well as the {$\alpha$}-enhancement in the 
cluster, thus providing vital clues to both the formation of globular 
clusters as well as time-scales for the formation of the Galaxy. 
The $\alpha$-enhancement is also important to know when deriving
an age from fitting of the main-sequence turn-off in the colour-magnitude
diagram as $\alpha$-enhancement changes the isochrones such that the 
age for a given metallicity gets younger (see Salasnich et al., 2000,
and for an application to metal-rich clusters Feltzing \& Johnson,
2002)

\smallskip
\noindent
{\bf Stellar sample}
The stars for the UVES observations were selected as being cluster 
members based on the individual stellar proper motions for stars in
NGC 6528 and the surrounding field (Feltzing \& Johnson, 2002). We 
selected both stars on the horizontal branch as well as on the upper red
giant branch.

\smallskip
\noindent
{\bf Stellar abundances}
The measured equivalent widths are analysed using a standard LTE
abundance calculation with MARCS model atmospheres 
(Gustafsson et al., 1975, and subsequent updates). Line data are
from Bensby et al. (2002 in prep.) - in particular Fe, Si and, Ti have
laboratory $\log gf$ while Ca and Cr have astrophysical $\log gf$'s.
  Comparing the difference between $\log gf$-values used by us and by 
Carretta et al. we find an  offset of -0.06 dex (our-carretta).

\begin{figure}  
\plottwo{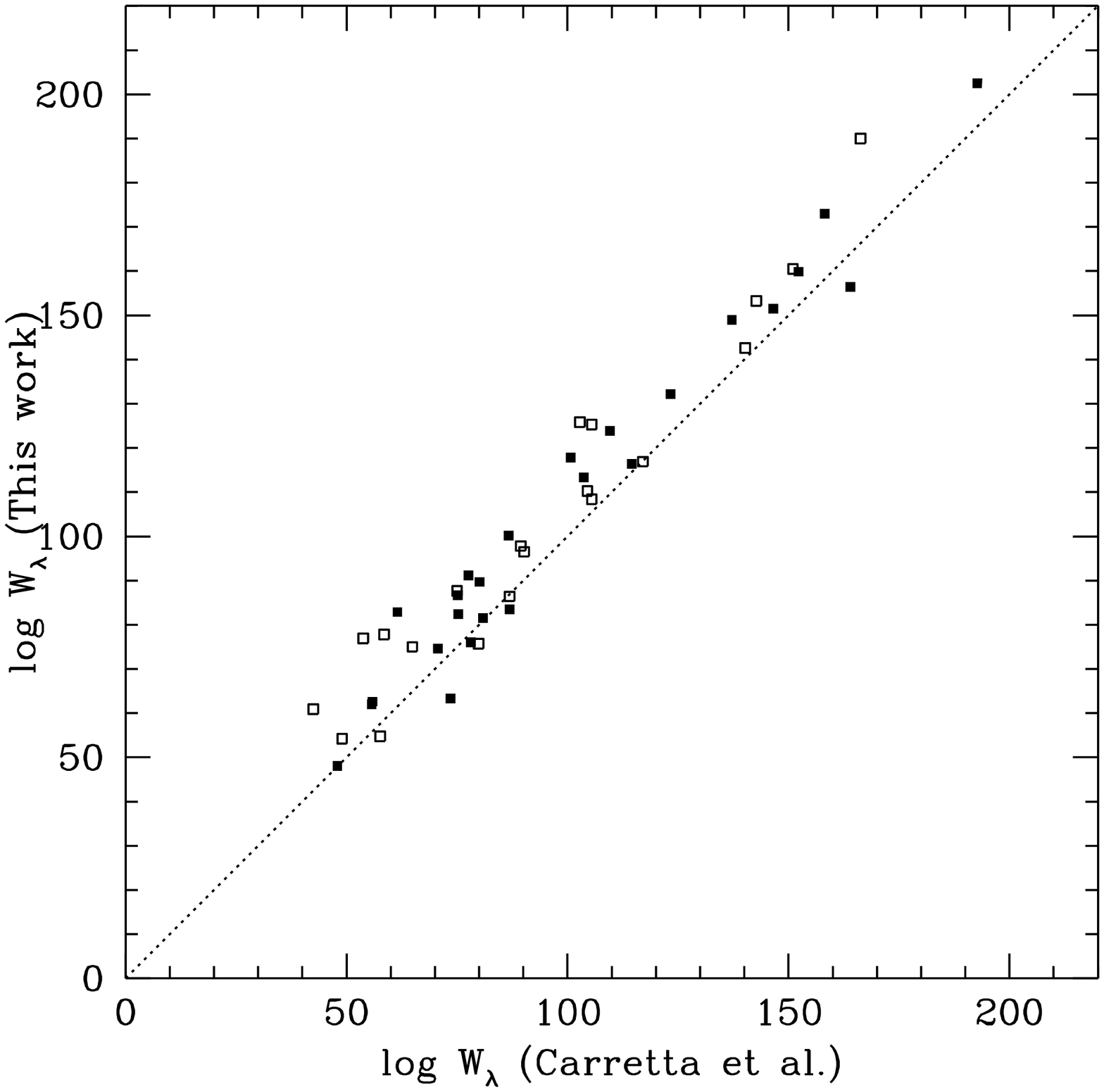}{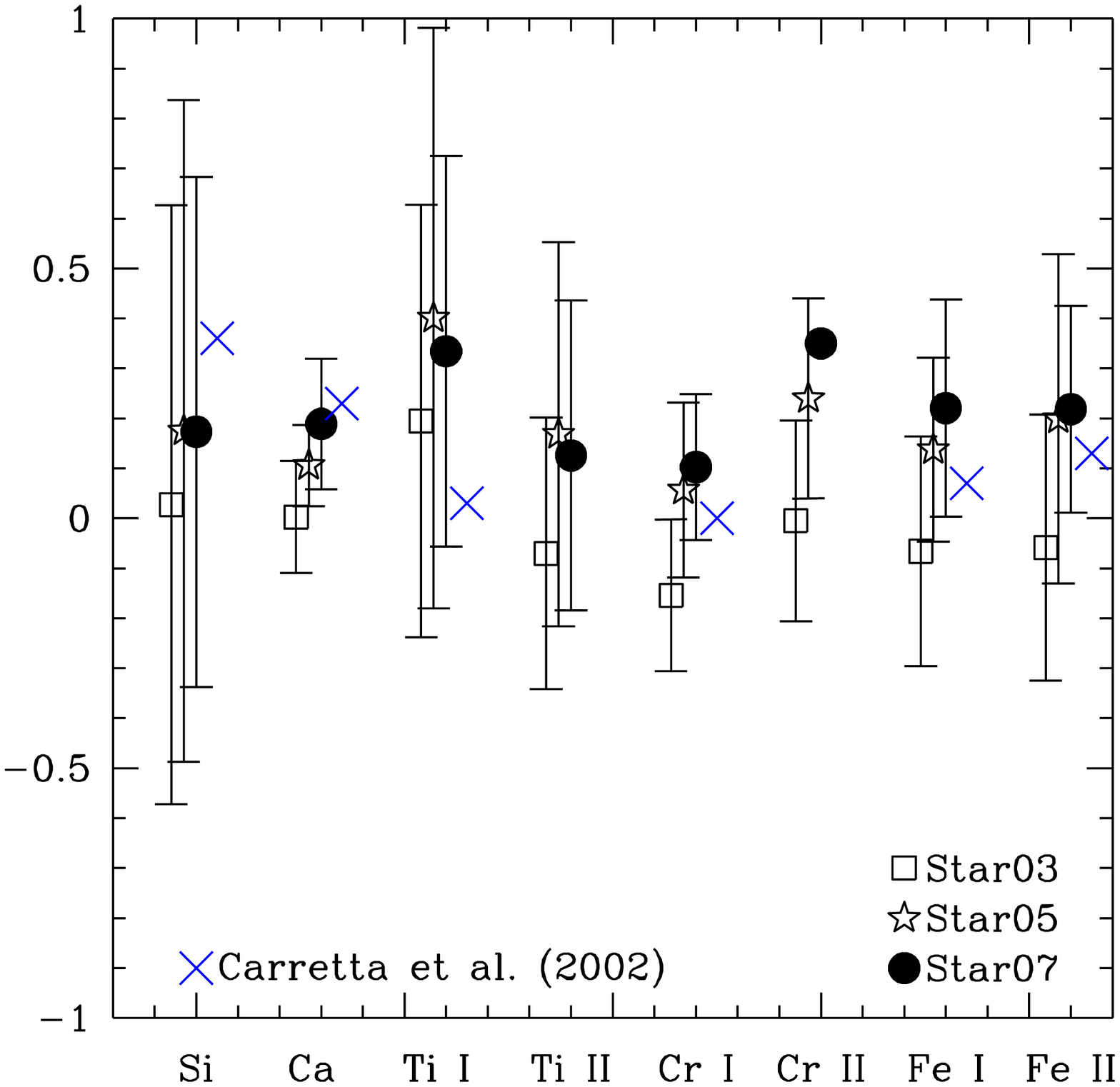}
\caption{Comparison of equivalent widths for lines in common between
us and  Carretta
et al.  for star 07 (their star 3025). A mean offset of 
8 m{\AA} is found.  This accounts for most of the difference
in our iron abundances.
In the right panel we show our, first,  mean abundances for the 
three stars analysed. The Carretta et al. abundances are their mean
abundances and the error-bars refer to the line-to-line scatter}
\end{figure}

\smallskip
\noindent
{\bf Discussion}
During this first analysis we found a number of 
issues that need to be tackled before the abundance analysis 
 can be finalised:\\
$\bullet~ $Definition of continuum -- as our comparison
of measured $W_{\lambda}$ for star 07  (star 
3025 in Carretta et al. 2001), Fig.1, shows the exact definition
of continuum is difficult in these metal-rich
giant stars. We are using synthetic spectra with appropriate
model parameters to gauge the continuum level. As stellar
parameters are iterated the continuum definition might need
redefining for certain wavelength regions.\\
$\bullet~ $Several other elements will require synthetic spectra
for abundance determinations. This is especially true for 
elements that have either very weak or very strong lines only 
(e.g. V and Mg). \\
$\bullet~ $ We will further investigate the large scatter for certain elements,
e.g. Si see Fig. 1. 
Also metal-rich field dwarf stars show a large line-to-line scatter 
(e.g. Feltzing \& Gustafsson 1998). Are these large scatters real?


\end{document}